\documentclass[reprint,superscriptaddress,showpacs,amsmath,amssymb,floatfix,aps,pra]{revtex4-1}

\usepackage{graphicx}
\usepackage{dcolumn}
\usepackage{bm}

\graphicspath{{./figures/}}

\DeclareMathOperator{\Tr}{Tr}

\usepackage{color}
\usepackage{braket}

\begin{document}


\title{Generation of one-million-mode continuous-variable cluster state \\ by unlimited time-domain multiplexing}



\author{Jun-ichi Yoshikawa}
\affiliation{Department of Applied Physics, School of Engineering, The University of Tokyo, 7-3-1 Hongo, Bunkyo-ku, Tokyo 113-8656, Japan}

\author{Shota Yokoyama}
\affiliation{Department of Applied Physics, School of Engineering, The University of Tokyo, 7-3-1 Hongo, Bunkyo-ku, Tokyo 113-8656, Japan}
\affiliation{Centre for Quantum Computation and Communication Technology, School of Engineering and Information Technology, University of New South Wales, Canberra, Australian Capital Territory 2600, Australia}

\author{Toshiyuki Kaji}
\affiliation{Department of Applied Physics, School of Engineering, The University of Tokyo, 7-3-1 Hongo, Bunkyo-ku, Tokyo 113-8656, Japan}

\author{Chanond Sornphiphatphong}
\affiliation{Department of Applied Physics, School of Engineering, The University of Tokyo, 7-3-1 Hongo, Bunkyo-ku, Tokyo 113-8656, Japan}

\author{\\Yu Shiozawa}
\affiliation{Department of Applied Physics, School of Engineering, The University of Tokyo, 7-3-1 Hongo, Bunkyo-ku, Tokyo 113-8656, Japan}

\author{Kenzo Makino}
\affiliation{Department of Applied Physics, School of Engineering, The University of Tokyo, 7-3-1 Hongo, Bunkyo-ku, Tokyo 113-8656, Japan}

\author{Akira Furusawa}
\email[]{akiraf@ap.t.u-tokyo.ac.jp}
\affiliation{Department of Applied Physics, School of Engineering, The University of Tokyo, 7-3-1 Hongo, Bunkyo-ku, Tokyo 113-8656, Japan}


\date{\today}

\begin{abstract}
In recent quantum optical continuous-variable experiments, the number of fully inseparable light modes has drastically increased by introducing a multiplexing scheme either in the time domain or in the frequency domain. 
Here, modifying the time-domain multiplexing experiment reported in Nature Photonics 7, 982 (2013), we demonstrate successive generation of fully inseparable light modes for more than one million modes. 
The resulting multi-mode state is useful as a dual-rail CV cluster state. 
We circumvent the previous problem of optical phase drifts, which has limited the number of fully inseparable light modes to around ten thousands, by continuous feedback control of the optical system. 
\end{abstract}


\maketitle


\section{\label{sec:intro}Introduction}

Quantum entanglement, which is often counterintuitive and was originally questioned by A.\ Einstein, B.\ Podolsky, and N.\ Rosen (EPR)~\cite{EPR}, is now recognized as versatile resource for quantum information protocols~\cite{Nielsen_Chuang,Furusawa_vanLoock}. 
In order to meet various requirements for various applications, an important technological development is to increase the number of available entangled qubits. 
In particular, measurement-based quantum computation (MBQC) requires the cluster type of a large-scale entangled state~\cite{Raussendorf.prl2001, Raussendorf.pra2003}, where individual components of the cluster state must be accessible by measuring devices. 
The number of entangled qubits with individual accessibility seems to be currently limited to a moderate scale~\cite{Ladd.nature2010, Monz.prl2011, Yao.nphoton2012}.

Shifting to the continuous variable (CV) counterpart~\cite{Zhang.pra2006, Menicucci.prl2006}, recently the number of fully inseparable qumodes of light fields drastically increased by introducing multiplexing either in the time domain~\cite{Menicucci.prl2010, Menicucci.pra2011, Yokoyama.nphoton2013} or in the frequency domain~\cite{Menicucci.pra2007, Menicucci.prl2008, Chen.prl2014, Roslund.nphoton2014}. 
The successes of the CV approach are based on the deterministic nature of field-quadrature squeezing. 
This is a strong advantage over photonic qubit-based approach, where photons or photons pairs are generated with low probabilities and rare success events are postselected~\cite{Yao.nphoton2012, Walther.nature2005}. 
On the flip side, the disadvantage of the CV MBQC has been considered as finite squeezing in resource cluster states, which results in accumulation of noises during computation~\cite{Ohliger.pra2010}. 
However, there has still been a possibility of a special encoding which can circumvent this problem, and indeed, a fault-tolerant threshold corresponding to about $-$20.5~dB of squeezing is recently proven by N.\ C.\ Menicucci~\cite{Menicucci.prl2014} for the Gottesman-Kitaev-Preskill (GKP) type of encoding~\cite{Gottesman.pra2001}. 
Although the threshold of the squeezing levels is still unreached, it is good news that CV cluster states with finite squeezing can be a sufficient resource for quantum computation.

The multiplexing allows many qumodes to be assigned to a single beam, which drastically reduces complexity and physical size of the optical system from the conventional implementation assigning a single qumode to a single optical beam~\cite{Yukawa.pra2008, Ukai.prl2011, Ukai.prl2011.2, Miwa.pra2010, Su.ncomm2013}. 
Although both of the time-domain multiplexing and the frequency-domain multiplexing would be important developmental directions, the time-domain multiplexing is advantageous because of time translation symmetry. 
With the time-domain multiplexing, a cluster state of qumodes can be an arbitrarily long chain in the longitudinal direction by extending the operating time of the cluster-state generator.

\begin{figure*}[tb]
\centering
\includegraphics[clip]{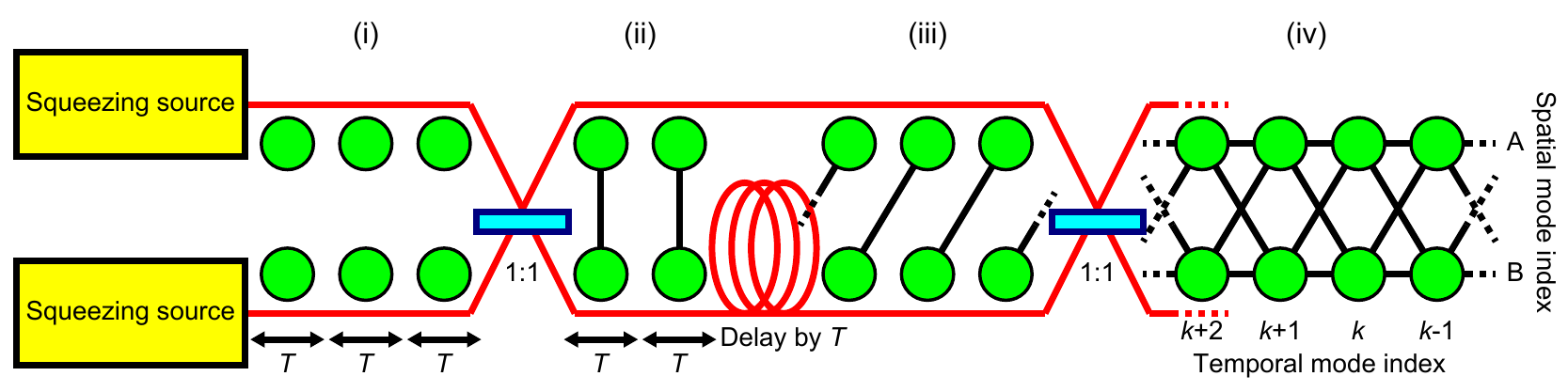}
\caption{
Schematic of the optical setup to generate a dual-rail CV cluster state multiplexed in the time domain. 
Accompanied is a graph image of the quantum state at each stage, where qumodes are represented by nodes and their correlations are represented by links.}
\label{fig:schematic}
\end{figure*}

In spite of the in-principle unlimitation of the time-domain-multiplexing, the previous experimental demonstration of a dual-rail CV cluster state was limited in the number of qumodes~\cite{Yokoyama.nphoton2013}, due to a technical reason as follows. 
The optical setup is a large Mach-Zehnder interferometer as depicted in Fig.~\ref{fig:schematic}, including a long optical delay line for a shift of qumodes on a rail. 
The mathematical derivation of the cluster state with this setup is contained in Appendix~\ref{sec:nullifier}. 
In order to operate this cluster-state generator, the relative phases at all interference points must be properly locked. 
For this purpose, modulated bright beams are injected into the optical paths of the cluster state as phase probes, and their classical interference signals are exploited as error signals for the feedback control. 
However, the modulated bright beams are noisy, which obscures objective quantum-level correlation signals of the cluster state. 
The noises were circumvented in the previous demonstration by chopping the bright beams and by detecting the cluster state when the bright beams are absent. 
Due to the above previous situation, the optical system was uncontrollable during the cluster-state generation. 
Phase drifts gradually accumulated and quantum correlations gradually degraded at the rear part of the cluster state. 
Finally, the full-inseparability criterion~\cite{vanLoock.pra2003} became unsatisfied after creation of around 16,000 qumodes within 1.3~ms~\cite{Yokoyama.nphoton2013}.

Here we demonstrate a new time-domain-multiplexing experiment, where a dual-rail CV cluster state with full inseparability is deterministically generated without limitation in the number of qumodes. 
The previous limitation explained above is gotten rid of by continuing the feedback control during the generation of a cluster state. 
Instead, the noises of the modulated bright beams are eliminated electrically after homodyne detections. 
The resulting effective squeezing levels are $-$4.3~dB, which are not as high as $-$5.0~dB for the first thousands of qumodes in the previous demonstration~\cite{Yokoyama.nphoton2013}, but in contrast the initial squeezing levels continue without degradation. 
The squeezing levels are sufficiently higher than $-$3.0 dB of the full-inseparability criterion~\cite{vanLoock.pra2003, Yokoyama.nphoton2013}. 
In principle, there is no limitation in the number of fully inseparable qumodes. 
We stopped the test of full inseparability after generation of about $1.2\times 10^6$ qumodes within 100 ms, simply because the data size of the homodyne signals reached 100~GB.

In Sec.~\ref{sec:theory}, we summarize theories regarding quantum correlations in the dual-rail CV cluster state, full inseparability criterion, and longitudinal mode functions. 
In Sec.~\ref{sec:method}, we describe experimental methods. 
In Sec.~\ref{sec:results}, we show experimental results and make discussions. 
In Sec.~\ref{sec:conclusion}, we summarize our results. 
For self-containedness, in Appendix~\ref{sec:nullifier}, we describe derivation of the quantum correlations in the dual-rail CV cluster state for an ideal case, and in Appendix~\ref{sec:inseparability}, we describe derivation of inequalities for full inseparability.

\section{\label{sec:theory}Theory}

\subsection{Quantum correlations in the cluster state}

The procedure of producing the dual-rail CV cluster state is as follows. 
First, we produce a series of equal-time two-mode squeezed states (ordinary EPR states) in two beams [(ii) of Fig.~\ref{fig:schematic}]. 
This can be done by combining two single-mode squeezed states at a balanced beamsplitter. 
The time-slot width of the multiplexing is denoted by $T$. 
Next, we add delay to half parts of the EPR states by the time-slot width $T$, by unbalancing the optical path lengths [(iii) of Fig.~\ref{fig:schematic}]. 
Finally, we combine the staggered EPR states by another balanced beamsplitter [(iv) of Fig.~\ref{fig:schematic}]. 
The resulting extended EPR (ExEPR) state is equivalent to the dual-rail CV cluster state up to local phase redefinitions~\cite{Yokoyama.nphoton2013}.

In Appendix~\ref{sec:nullifier}, the above procedure is mathematically followed for the ideal, infinitely squeezed case. 
A convenient way of representing the state, utilized in Appendix~\ref{sec:nullifier}, is to specify the state by its nullifiers~\cite{Gu.pra2009} in the form of a linear combination of single-mode quadrature operators $\hat{x}_{\Lambda,k}$ and $\hat{p}_{\Lambda,k}$. 
Here, $\Lambda\in\{A,B\}$ denotes the spatial mode index, and $k \in \mathbb{N}$ denotes the temporal mode index, as depicted in (iv) of Fig.~\ref{fig:schematic}. 
Quadrature operators obey a commutation relation similar to that of position and momentum operators, $[\hat{x}_{\Lambda,k},\hat{p}_{\Lambda^\prime,k^\prime}] = i\hbar\delta_{\Lambda\Lambda^\prime}\delta_{kk^\prime}$, where $\delta_{qq^\prime}$ is the Kronecker delta. 
Appendix~\ref{sec:nullifier} is summarized in the following set of nullifiers, 
\begin{subequations}
\begin{align}
\hat{X}_k &:= \hat{x}_{A,k}+\hat{x}_{B,k}+\hat{x}_{A,k+1}-\hat{x}_{B,k+1}, \\ 
\hat{P}_k &:= \hat{p}_{A,k}+\hat{p}_{B,k}-\hat{p}_{A,k+1}+\hat{p}_{B,k+1}, 
\end{align}
\end{subequations}
The ideal ExEPR state is specified by the set of these nullifiers via the relations $\hat{X}_k\ket{\text{ExEPR}}=0$ and $\hat{P}_k\ket{\text{ExEPR}}=0$ for all $k$. 
These nullifiers also express the quantum correlations, which are experimentally observable. 
Note that all the nullifiers are mutually commutable, thus giving a simultaneous eigenstate.

In reality, the resource squeezing levels are finite. 
Therefore, when the nullifiers $\hat{X}_k$ and $\hat{P}_k$ are measured with respect to an experimental ExEPR state, there are residual noise variances, 
\begin{align}
\braket{\hat{X}_k^2} &= 2\hbar e^{-2r_{x,k}}, &
\braket{\hat{P}_k^2} &= 2\hbar e^{-2r_{p,k}}. 
\label{eq:nullifier_var}
\end{align}
Here, the bracket $\braket{\hat{O}}$ represents the mean value of an observable $\hat{O}$. 
The squeezing parameters $r_{x,k}$ and $r_{p,k}$ correspond to those of the first and second single-mode squeezing sources in (i) of Fig.~\ref{fig:schematic}, respectively, when the subsequent transformations and measurements are ideal, which is apparent from calculations in Appendix~\ref{sec:nullifier}. 

Then, a natural question is what levels of squeezing are required for the entanglement. 
Here we resort to the van Loock-Furusawa criterion for full inseparability~\cite{vanLoock.pra2003, Yokoyama.nphoton2013}. 
A sufficient condition for full inseparability is, 
\begin{align}
\braket{\hat{X}_k^2} &< \hbar, & \braket{\hat{P}_k^2} &< \hbar, & \text{for all $k$}. 
\end{align}
This condition corresponds to $-$3.0 dB squeezing of each nullifier. 
The derivation is contained in Appendix~\ref{sec:inseparability}.

\subsection{\label{ssec:mode}Longitudinal mode functions}

In our demonstration, we use continuous-wave (CW) local oscillators (LOs) for homodyne detections. 
Therefore, the homodyne detection system continuously outputs real values $\hat{x}_{\Lambda}^\text{det}(t)$ or $\hat{p}_{\Lambda}^\text{det}(t)$ with $\Lambda\in\{A,B\}$, which correspond to instantaneous quadrature operators $\hat{x}_{\Lambda}(t)$ or $\hat{p}_{\Lambda}(t)$ but filtered by the detection system with finite bandwidth.

The quadrature operator $\hat{x}_{\Lambda,k}$ of each qumode, specified by a longitudinal mode function $f_k(t)=f_0(t-kT)$ which should be contained in the $k$-th time slot $kT\le t\le (k+1)T$, is connected to the instantaneous quadrature operators $\hat{x}_{\Lambda}(t)$ via the relation, 
\begin{align}
\hat{x}_{\Lambda,k} = \int dt f_k(t)\hat{x}_{\Lambda}(t) = \int dt f_0(t)\hat{x}_{\Lambda}(t+kT). 
\label{eq:integral}
\end{align}
However, what we can directly choose in the experiment is not the optical longitudinal mode function $f_k(t)$ but a computational weight function $g_k(t)$, with which we perform weighted integration, 
\begin{align}
\hat{x}_{\Lambda,k} = \int dt g_k(t)\hat{x}_{\Lambda}^\text{det}(t) = \int dt g_0(t)\hat{x}_{\Lambda}^\text{det}(t+kT). 
\end{align}
This integration is totally a postprocessing after acquisition of full-time quadrature values $\hat{x}_{\Lambda}^\text{det}(t)$. 
The situation is the same for the conjugate quadrature operator $\hat{p}_{\Lambda,k}$. 
Note that the integration is actually replaced by a summation in accordance with the sampling rate of data acquisition.

If the response of the homodyne detection system is sufficiently flat, we can identify $g_k(t)$ with $f_k(t)$. 
However, this is not the case in this demonstration, in contrast to the previous demonstration~\cite{Yokoyama.nphoton2013}, because of the electric filters applied to the homodyne signals for the reasons explained in Sec.~\ref{sec:intro} and in more detail in Sec.~\ref{ssec:modulation}. 
They are connected as $g_k(t)\propto(e\ast f_k)(t)$ via the impulse response function $e(t)$ of the detection system including the electric filters, where $\ast$ denotes convolution. 
Even if we choose the weight function $g_k(t)$ so that they are nonzero only in the time slot $kT\le t\le (k+1)T$, by which the weight functions themselves are orthonormal $\int dt g_k(t)g_{k^\prime}(t) = \delta_{kk^\prime}$, the orthonormality of actual optical modes $\int dt f_k(t)f_{k^\prime}(t) = \delta_{kk^\prime}$ is not guaranteed in general.

\begin{figure}[tb]
\centering
\includegraphics[clip]{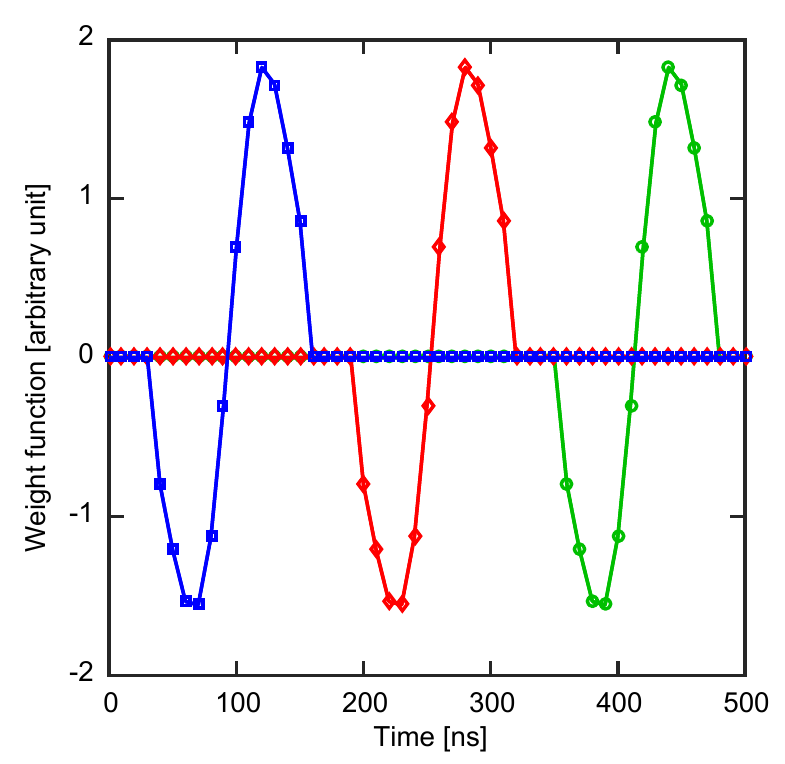}
\caption{
First three weight functions $g_0(t)$, $g_1(t)$, $g_2(t)$, marked by squares, diamonds, and circles, respectively. 
They are discretized at intervals of 10~ns, in accordance with the sampling rate 100~MHz of data acquisition.
}
\label{fig:weightfunc}
\end{figure}

The filter is a combination of both high-pass and low-pass filters, and the parameters are described in Sec.~\ref{ssec:modulation}. 
A filter delays signals in general. 
Therefore, when the response $e(t)$ is deconvoluted from the weight function $g_k(t)$, the optical mode function $f_k(t)$ is stretched in the forward direction, which may ruin the mode independence to some extent. 
Especially, the effect of a high-pass filter is not negligible in our setup, which produces a long tail in the forward direction. 
In order to suppress the nonorthogonality of the actual optical modes, a possible method is to use a weight function localized to the latter part of the time slot. 
This is a workable method, but is not sufficient in our case, as will be discussed by using autocorrelations of shot noises in Sec.~\ref{ssec:autocorr}.

\begin{figure*}[tb]
\centering
\includegraphics[clip]{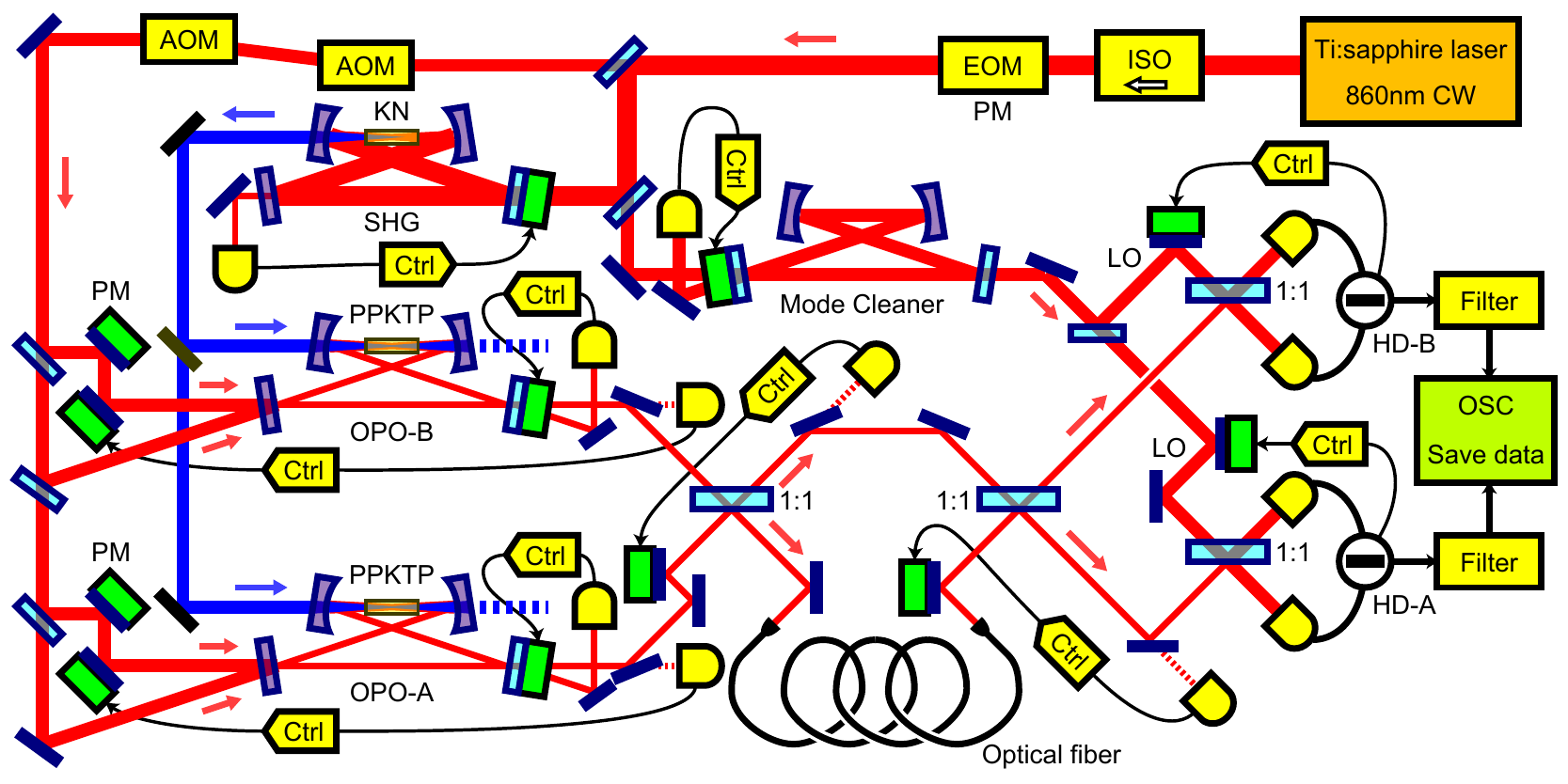}
\caption{Experimental setup. ISO, optical isolator; PM, phase modulation; HD, homodyne detector; Osc, oscilloscope; Ctrl, feedback controller.}
\label{fig:setup}
\end{figure*}

Therefore, here we use a weight function which has both positive and negative parts. 
By adjusting the balance between the positive part and the negative part, the long tail produced by the high pass filter is canceled. 
The actual unnormalized weight functions utilized in the experimental data analysis are shown in Fig.~\ref{fig:weightfunc}. 
The unnormalized weight functions are the following functions, 
\begin{align}
g_k(t)=
\begin{cases}
\exp[-\gamma^2(t-t_k)^2] & \\
\quad\times(t-t_k+t_c), & \text{if $2|t-t_k| \le T_w$}, \\
0, & \text{otherwise}, 
\end{cases}
\label{eq:weightfunc}
\end{align}
discretized at intervals of 10~ns. 
The width of the weight function window $T_w$ is 120~ns, and the envelope fall time $1/\gamma$ is 40~ns. 
The balance between the positive part and the negative part is adjusted through an optimization parameter $t_c$, which is determined to be 2~ns. 
The center of the weight function is $t_k=t_0+kT$, where $k \in \mathbb{N}$ is the temporal index as usual. 
The mode interval $T$ is 160~ns, and $t_0 = 95$~ns in Fig.~\ref{fig:weightfunc}. 
Note that we do not have to care about the normalization of the weight functions in the actual data analysis, because the obtained quadrature values are normalized by using corresponding shot noises acquired in the same conditions.

On the other side of the coin, the wavy weight functions lead to increased contribution of higher frequency components, compared to the previous demonstration~\cite{Yokoyama.nphoton2013}. 
However, the squeezing level degrades at higher frequencies with our squeezing sources, due to a finite bandwidth of optical cavities. 
The resulting correlation levels of $-$4.3~dB is not as high as the previous $-$5.0~dB for this reason.

\section{\label{sec:method}Experimental Methods}

\subsection{Optical setup}

Figure~\ref{fig:setup} shows the experimental setup. 
The light source is a continuous-wave (CW) Ti:sapphire laser (SolsTiS, M Squared Lasers) operating at the wavelength of 860~nm and the power of about 2~W. 
About 1~W of the laser power is sent to a second harmonic generation (SHG) cavity. 
It is a bow-tie shaped cavity with a round-trip length of 500~mm, containing a KNbO$_3$ (KN) crystal as a nonlinear optical medium. 
The generated CW second harmonic light is used as pump beams for two optical parametric oscillators (OPOs) operating below threshold. 
The power of the pump beam for each OPO is about 150~mW. 
Each OPO is a bow-tie shaped cavity with a round-trip length of 230~mm, containing a periodically-poled KTiOPO$_4$ (PPKTP) crystal (Raicol Crystals) as a nonlinear optical medium. 
It generates a squeezed light beam, which is treated as a stream of single-mode squeezed vacuum states. 
The bandwidth of the squeezing basically corresponds to the bandwidth of the OPO cavity, whose half width at half maximum is 17~MHz.

The two squeezed light beams from the OPOs are sent to an asymmetric Mach-Zehnder interferometer, composed of two balanced beamsplitters and an optical delay line. 
The delay line is in one of the two arms of the Mach-Zehnder interferometer, and implemented with an optical fiber. 
The length of the optical fiber is about 30~m, from which the time slot width $T$ of the multiplexing is about 160~ns. 
Both ends of the fiber are anti-reflection coated, and the coupling efficiency is maximized by using special fiber aligners (FA1000S, First Mechanical Design). 
Total amount of losses regarding the optical fiber, including the coupling inefficiency and the propagation losses, was about 11\%.

In order to characterize the resulting ExEPR state, the two output beams are each subject to a balanced homodyne detection. 
The bandwidth of the homodyne detectors without the electric filter is more than 30~MHz. 
The power of a CW local oscillator (LO) beam for each homodyne detection is 10~mW. 
The transverse mode of the LO beams is cleaned and stabilized by a cavity. 
The interference visibility between the ExEPR beams and the LO beams were about 97\% on average. 
The quantum efficiency of the homodyne photodiodes is about 99\%.

\subsection{\label{ssec:modulation}Modulations, filters, and data acquisition}

In order to probe the optical phases and thereby to lock the relative phases of all optical interferences by feedback control, modulated bright probe beams are utilized. 
At each interference point, there is a corresponding photo-detector, and demodulation of the detector signal gives an error signal, which is fed back to a piezoelectrically actuated mirror, as depicted in Fig.~\ref{fig:setup}. 
First, a phase modulation at 32~MHz is commonly added by an electro-optic modulator (EOM) just after the Ti:sapphire laser output. 
This modulation is utilized for the Pound-Drever-Hall lock~\cite{Black.ajp2001} of the optical cavities including the OPOs. 
Next, two probe beams are phase modulated by piezoelectric transducers (PZTs) at 231~kHz and 326~kHz, respectively, and then injected into the individual OPOs from the back side of a highly reflective mirror composing the OPO. 
The powers of the probe beams are adjusted to about 2.5~$\mu$W at the output of the OPOs. 
These lower-frequency modulations are utilized for the lock of the following three. 
Firstly, the phase-sensitive parametric amplification at each OPO is locked to the point minimizing the probe beam power, by which the phase of the squeezed light is associated with that of the probe beam. 
Secondly, the two interferences of the Mach-Zehnder interferometer are locked, where the beat signal of the two modulations is exploited by demodulating at the difference frequency of 95~kHz. 
Thirdly, the LO phases are locked so that the homodyne detectors measure either the objective $\hat{x}_{\Lambda}^\text{det}(t)$ or $\hat{p}_{\Lambda}^\text{det}(t)$ quadrature values. 
The measurement basis is selected via the choice of the demodulation frequency between 231~kHz or 326~kHz. 
These modulations of bright probe beams become unwanted noises in the homodyne detections, which wipe out the objective quantum correlations and thus must be removed somehow.

As stated in Sec.~\ref{sec:intro}, these noises were eliminated at the level of the optical system in the previous work~\cite{Yokoyama.nphoton2013}, employing periodic switching. 
The probe beams were alternately turned on and off by using a pair of acousto-optic modulators (AOMs), and the feedback control of the optical system was active when the probe beams were on, while the optical system was held and the ExEPR state was tested when the probe beams were off. 
In that demonstration, uncontrollable drifts of the optical system were negligible for first several thousands of qumodes, but gradually degraded the correlation in the ExEPR state, finally spoiling the full inseparability after creation of about 16,000 qumodes. 
In contrast, here we employ the method of eliminating the noises electrically after the homodyne detections, by using frequency filters. 
In this demonstration, the switching AOMs are always on, and the feedback control of the optical system continues during creation of the ExEPR state.

\begin{figure}[tb]
\centering
\includegraphics[clip]{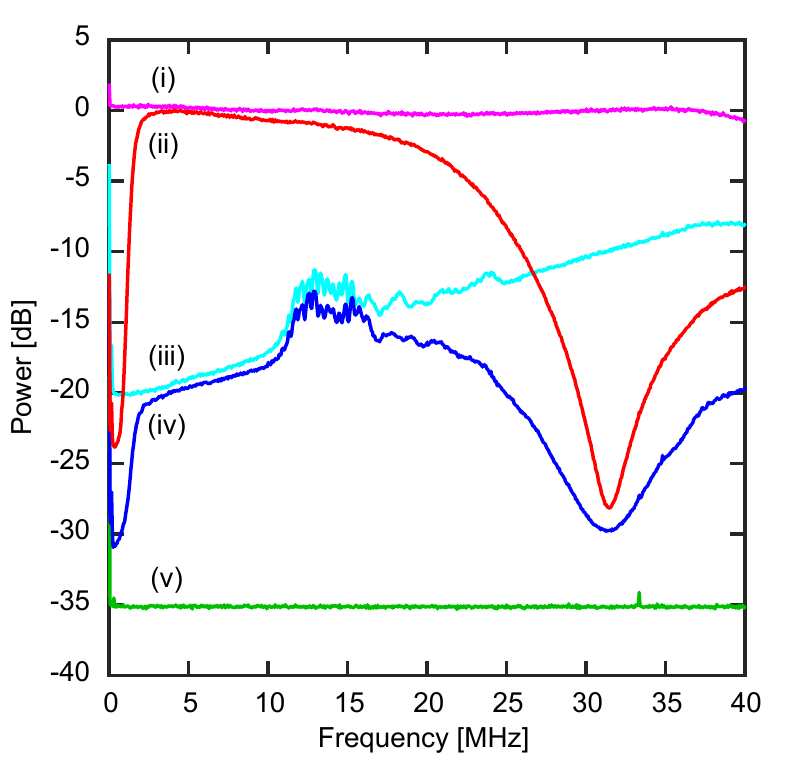}
\caption{
Power spectra with and without the electric filter. 
(i) Optical shot noise without the electric filter. 
(ii) Optical shot noise with the electric filter. 
(iii) Dark noise without the electric filter. 
(iv) Dark noise with the electric filter. 
(v) Noise floor of the oscilloscope. 
The dark noise is the electric noise of the detection system when the LO beam is absent. 
2048 data points are used for each fast Fourier transform, and the resulting power spectrum is averaged over 6000 times. 
}
\label{fig:fft_noise}
\end{figure}

\begin{figure*}
\centering
\includegraphics[clip]{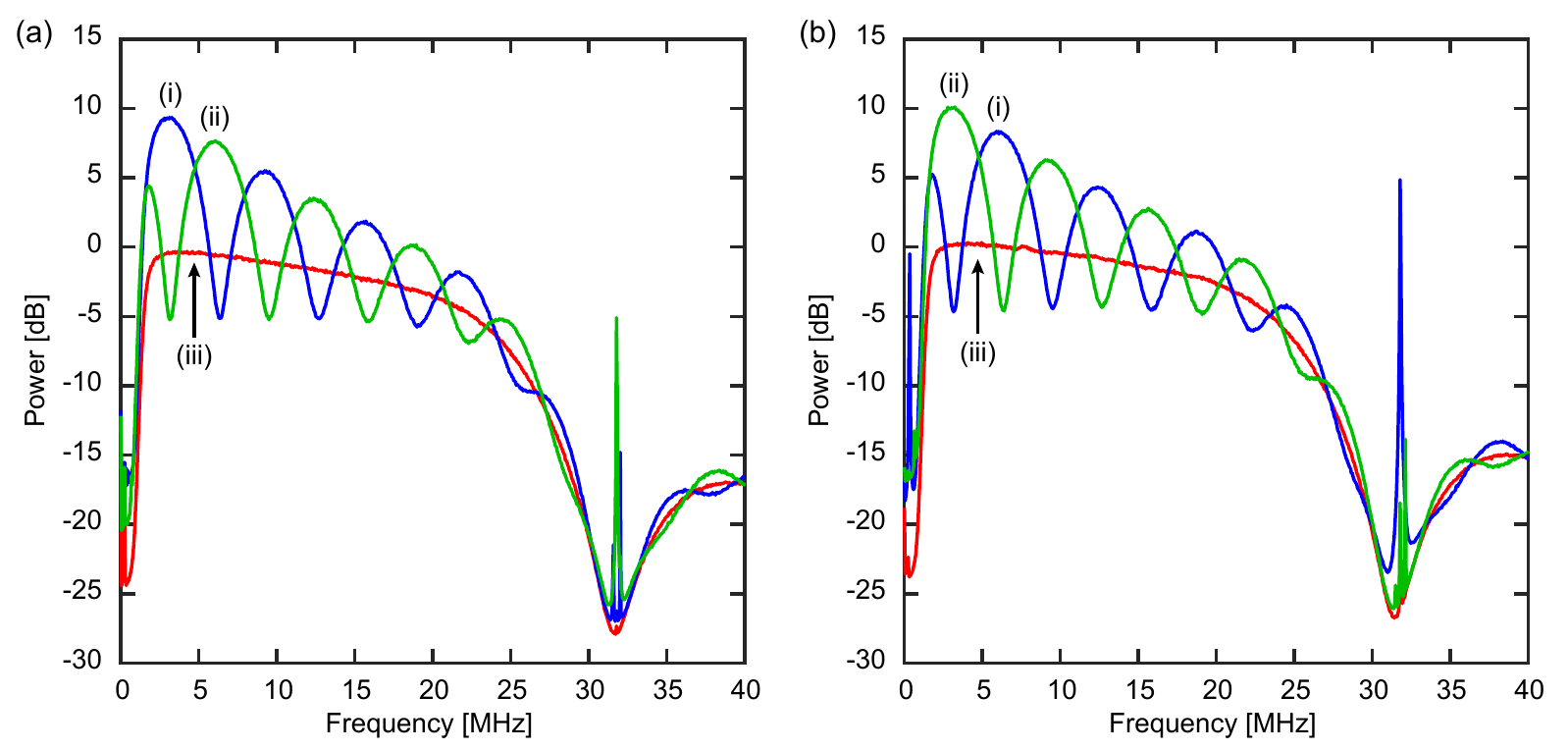}
\caption{
Power spectra of homodyne detection signals. 
(a) HD-A. (b) HD-B. 
(i) $\hat{x}$ quadrature signals of the ExEPR output. 
(ii) $\hat{p}$ quadrature signals of the ExEPR output. 
(iii) Optical shot noises for reference. 
2048 data points are used for each fast Fourier transform, and the resulting power spectrum is averaged over 6000 times. 
}
\label{fig:fft_quad}
\end{figure*}

The electric filter applied to each homodyne signal in this demonstration is a serial combination of a 5th-order high pass filter with the cutoff frequency of 1.5~MHz, a 3rd-order inverse Chebyshev low-pass filter which has a notch characteristic at 32~MHz, and a 3rd-order low-pass filters with the cutoff frequency of 40~MHz for the purpose of anti-aliasing. 
After the electric filtering and appropriate low-noise amplification, the homodyne signals are digitized by an 8-bit oscilloscope (DPO7054, Tektronix) with the sampling rate of 100~MHz. 
The data acquisition frame width is 100~ms, and we acquire 1500 frames of data for each measurement quadrature in order to determine the residual noise variance of each nullifier with acceptable precision. 
In the same condition, we also acquire optical shot noises for references, which correspond to vacuum fluctuations. 
The shot noises are obtained by simply blocking the optical beam paths of the ExEPR dual rail.

Figure~\ref{fig:fft_noise} shows the power spectra from the detection system with and without the electric filters, obtained from the fast Fourier transform (FFT) of oscilloscope data. 
We can see that the signals around the modulation frequencies are largely attenuated by the filters.

\section{\label{sec:results}Results and Discussion}

\subsection{Power spectra of homodyne signals}

Since now the ExEPR state is stabilized by the continuous feedback, first we discuss the power spectra of the homodyne signals. 
Figure~\ref{fig:fft_quad} shows the power spectra of the quadrature signals $\hat{x}_{\Lambda}^\text{det}(t)$ and $\hat{p}_{\Lambda}^\text{det}(t)$ of the output ExEPR beams, together with that of the optical shot noise $\hat{x}_{\Lambda}^\text{vac,det}(t)$ for references, for each of the two output ports $\Lambda\in\{A,B\}$. 
We can see that the entity of the ExEPR state generator output is, in the frequency domain, an oscillation between $\hat{x}$-squeezed states and $\hat{p}$-squeezed states, and the squeezed quadratures are opposite between the two output ports. 
This interesting point was not discussed in the previous work~\cite{Yokoyama.nphoton2013}.

By using the schematic in Fig.~\ref{fig:schematic}, we discuss how this periodic behavior in the frequency domain is understood. 
The situation is that two squeezed beams, one is $\hat{x}$-squeezed and the other is $\hat{p}$-squeezed, enter a Mach-Zehnder interferometer with asymmetric lengths of arms. 
When we focus on a single sideband frequency component, the unbalance of the arm lengths acts as just a phase shift. 
The amount of the phase shift is $\theta=2\pi fT$, proportional to the sideband frequency $f$ and the difference between the propagation times of the two arms $T$. 
Then, the two-input two-output Mach-Zehnder interferometer as a whole works as a single beamsplitter with a reflectivity $R=\cos\theta$ dependent on the phase shift $\theta$. 
At the phases of $R=1$, the $\hat{x}$-squeezed beam fully goes to one output port and the $\hat{p}$-squeezed beam the other output port, and at the phases of $R=0$, the opposite occurs. 
Between the above two cases, the beamsplitter combines the two squeezed beams and makes entanglement between the output ports. 
These cases periodically appear in the frequency domain because the phase shift $\theta$ is proportional to the sideband frequency $f$. 
The period of the oscillation corresponds to the propagation time difference $T$.

\subsection{\label{ssec:autocorr}Orthogonality of qumodes}

\begin{figure*}
\centering
\includegraphics[clip]{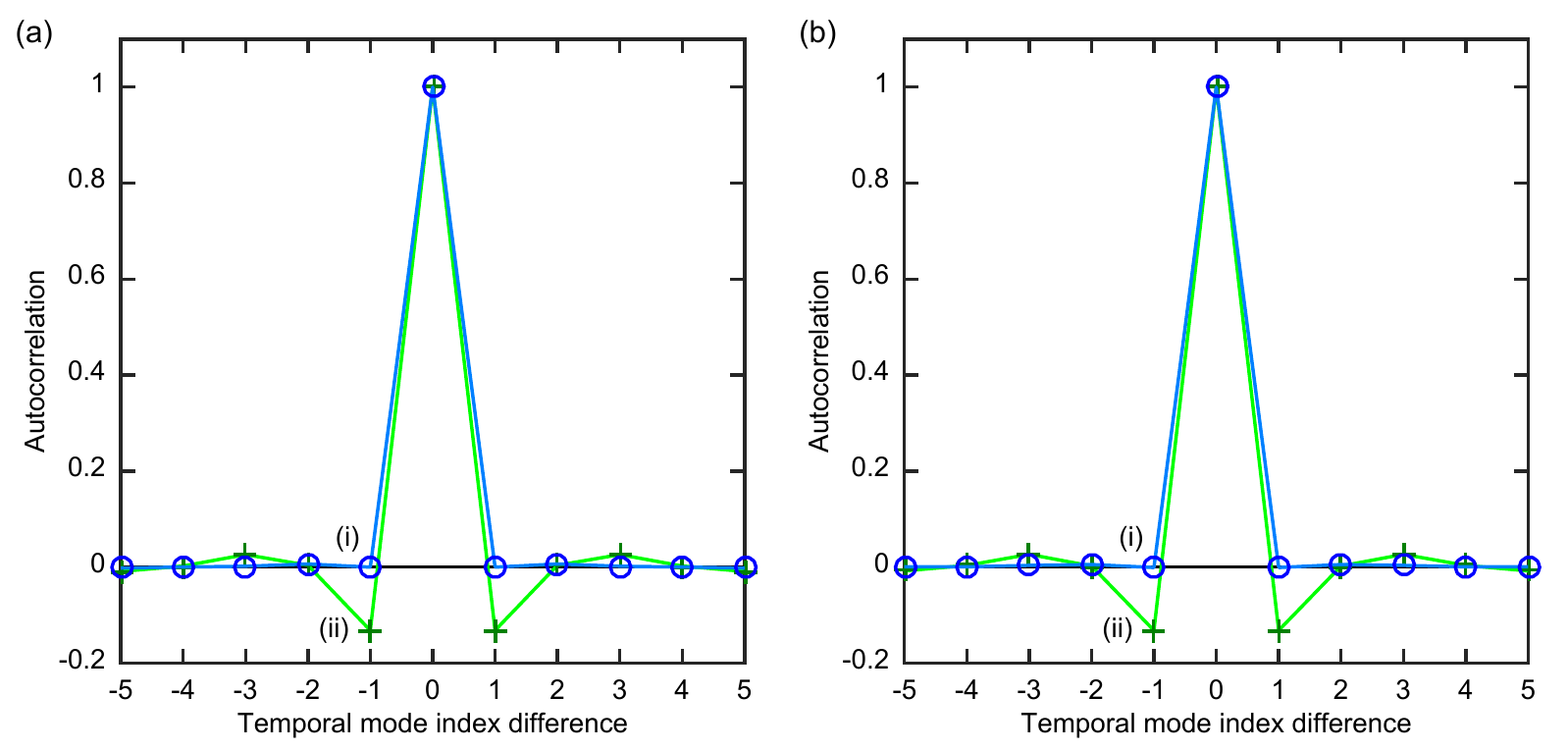}
\caption{
Autocorrelation of a series of integrated shot noises. 
(a) HD-A. (b) HD-B. 
(i) When the weight functions in Eq.~\eqref{eq:weightfunc} and Fig.~\ref{fig:weightfunc} are used. 
(ii) When only the positive part of the weight functions are used. 
Average is taken over 1.5$\times10^6$ times. 
}
\label{fig:corr}
\end{figure*}

We apply the weighted integration to the above homodyne signals by using weight functions described in Sec.~\ref{ssec:mode}. 
Before that, we check the orthogonality of the qumodes by applying the weighted integration to the optical shot noises, 
\begin{align}
\hat{x}_{\Lambda,k}^\text{vac} 
= \int dt g_k(t)\hat{x}_{\Lambda}^\text{vac,det}(t) 
= \int dt f_k(t)\hat{x}_{\Lambda}^\text{vac}(t). 
\end{align}
An unfiltered shot noises $\hat{x}_{\Lambda}^\text{vac}(t)$ is a white noise whose autocorrelation is represented by a Dirac delta function. 
Therefore, if the integrated shot noise values of a temporal mode $\hat{x}_{\Lambda,k}^\text{vac}$ have some correlation with those of neighboring temporal modes $\hat{x}_{\Lambda,k+m}^\text{vac}$, $m\neq0$, this correlation is due to the overlap between the optical mode functions $f_k(t)$ and $f_{k+m}(t)$. 

Figure~\ref{fig:corr} shows the correlation of the integrated shot noise values among neighboring qumodes, 
\begin{align}
C_{\Lambda}(m) := \frac{\langle\hat{x}_{\Lambda,k}^\text{vac}\hat{x}_{\Lambda,k+m}^\text{vac}\rangle}{\langle(\hat{x}_{\Lambda,k}^\text{vac})^2\rangle}, 
\end{align}
for $|m|\le 5$. 
As we can see with trace (i), the correlation is almost canceled, $|C_{\Lambda}(m)|<0.01$ for $m\neq0$, when the weight functions $g_k(t)$ in Eq.~\eqref{eq:weightfunc} and Fig.~\ref{fig:weightfunc} is used. 
For comparison, trace (ii) shows the correlation when only the positive parts of the weight functions $g_k^+(t)\propto\max(g_k(t),0)$ are used. 
In spite of the spacing between neighboring weight functions, a negative correlation $C_\Lambda(\pm1)\approx-0.13$ is observed, which is mainly due to the high-pass filter. 
This correlation is canceled by the negative parts of the weight functions. 
An insight regarding this cancellation is that the wavy weight functions make little use of the low-frequency components, which are affected by the high-pass filter and cause overlap between qumodes.

Note that the fractions of a qumode invading neighboring qumodes correspond to square of the correlation coefficients $|C_{\Lambda}(m)|^2$, which are very small and thus experimentally negligible.

\subsection{Full inseparability}

\begin{figure*}[t]
\centering
\includegraphics[clip]{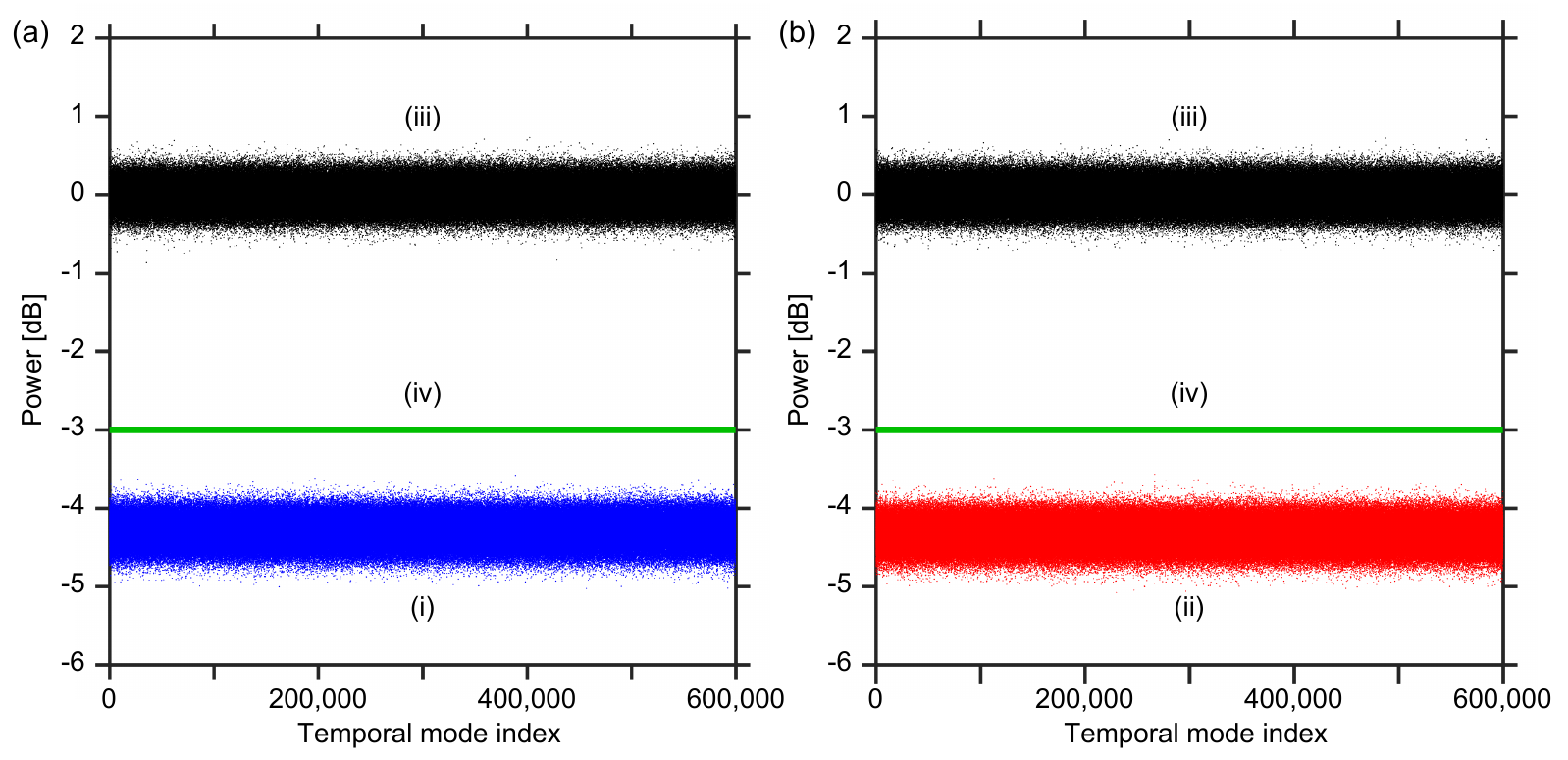}
\caption{
Residual noise variances of the nullifiers. 
(i) Variances of $\hat{X}_k$. 
(ii) Variances of $\hat{P}_k$. 
(iii) Corresponding reference shot noises. 
(iv) A sufficient condition of $-$3~dB bound for full inseparability. 
Horizontal axes are the temporal mode index $k$. 
Any correction of optical losses or detection noises is not applied. 
}
\label{fig:variance}
\end{figure*}

Finally we check the inseparability condition. 
Figure~\ref{fig:variance} shows the residual variances of the nullifiers, $\braket{\hat{X}_k^2}$ and $\braket{\hat{P}_k^2}$ in Eq.~\eqref{eq:nullifier_var}, together with the variances of the same observables calculated from the shot noises for references. 
We can clearly see the variances well squeezed below the $-$3~dB bound for full inseparability, for up to $k=6\times10^5$. 
Taking into account the two spatial mode indices $A$ and $B$, the number of qumodes tested here is $1.2\times10^6$. 
As mentioned in Sec.~\ref{ssec:modulation}, the variance of each nullifier is averaged over 1500 times. 
The average and the standard deviation of the squeezing levels over the $6\times10^5$ temporal indices is $-$4.3$\pm$0.2~dB for $\hat{X}_k$ and $-$4.3$\pm$0.2~dB for $\hat{P}_k$. 
The worst squeezing level among the $6\times10^5$ temporal indices is $-$3.6~dB for $\hat{X}_k$ and $-$3.5~dB for $\hat{P}_k$. 
They will approach the average $-$4.3~dB if we could increase the number of data frames for the averaging.

In contrast to the previous work~\cite{Yokoyama.nphoton2013}, the squeezing levels do not degrade during the generation of over-one-million qumodes. 
This is owing to the stabilized optical system by introducing the continuous feedback.

\section{\label{sec:conclusion}Conclusion}

We have demonstrated the deterministic generation and verification of a fully inseparable dual-rail CV cluster state consisting of more than one million qumodes of light, by employing a time-domain multiplexing scheme. 
Compared to the previous work~\cite{Yokoyama.nphoton2013}, the cluster state generator does not degrade during operation, owing to the continuous feedback of the optical system. 
We can in principle further increase the number of qumodes, but we stopped at around one million qumodes because of the data size for verification. 
Time domain multiplexing is one of the key technologies of CV information processing. 
Open questions from technological aspects are how to combine the cluster states with photon counters aiming at universal processing~\cite{Gu.pra2009}, as well as how to increase the squeezing levels toward fault tolerance. 
We mention recent experiments of optical filters for quantum states~\cite{Takeda.pra2012}, as well as real-time integration of homodyne signals~\cite{Ogawa.prl2016}, which in the future may be combined with this entanglement multiplexing technologies. 


\section*{Acknowledgment}

This work was partly supported by 
PDIS and APSA commissioned by the MEXT, 
JSPS KAKENHI, 
CREST of the JST, 
and the SCOPE program of the MIC, 
of Japan, 
and ARC CQC2T (No.\ CE1101027) of Australia. 
S.\ Y.\ acknowledges support from ALPS, and JSPS Postdoctoral Fellowships for Research Abroad, of Japan. 

\appendix
\section{\label{sec:nullifier}Nullifier transformation}

The procedure of creating the dual-rail CV cluster state multiplexed in the time domain, depicted in Fig.~\ref{fig:schematic}, can be understood by following the transformation of the nullifiers for the ideal, infinitely squeezed case. 
A nullifier is an operator $\hat{N}$ which nullifies a specific state vector $\hat{N}\ket{\psi}=0$. 
Any state vector can be completely specified by a set of nullifiers, which corresponds to the Lie algebra of the CV stabilizer group~\cite{Gu.pra2009}. 
It is often helpful to use them to specify a quantum state instead of directly expressing a state vector in a superposition form.

For the ideal cases of infinite squeezing, the nullifiers of the ExEPR state can be Hermitian, linear combinations of quadrature operators $\{\hat{x}_{\Lambda,k},\hat{p}_{\Lambda,k}:\Lambda\in\{A,B\},k\in\mathbb{N}\}$, which are measurable in the experiment. 
In the following, we discuss the nullifier transformations, starting from a number of single-mode squeezed states.

We suppose that the squeezing source $A$ produces a series of single-mode squeezed states whose $\hat{x}$ quadrature is squeezed, while the squeezing source $B$ produces a series of single-mode squeezed states whose $\hat{p}$ quadrature is squeezed. 
In the limit of infinite squeezing levels, the initial squeezed states approach quadrature eigenstates $\ket{x=0}_{A,k}$ and $\ket{p=0}_{B,k}$. 
The whole multiplexed initial state, which is a tensor product of the squeezed states 
\begin{align}
\ket{\text{MSQZ}}:=\bigotimes_{k\in\mathbb{N}}\ket{x=0}_{A,k}\ket{p=0}_{B,k}, 
\end{align}
is uniquely specified by a set of its nullifiers $\{\hat{x}_{A,k},\hat{p}_{B,k}:k\in\mathbb{N}\}$ with the following nullification relation, 
\begin{subequations}
\begin{align}
\hat{x}_{A,k}\ket{\text{MSQZ}} &= 0, \\
\hat{p}_{B,k}\ket{\text{MSQZ}} &= 0.
\end{align}
\end{subequations}

Next, a beamsplitter interference $\hat{U}_\text{BS}$ unitarily transforms the initial multiplexed squeezed state into the multiplexed EPR state as 
\begin{align}
\ket{\text{MSQZ}} \to \ket{\text{MEPR}} :=& \hat{U}_\text{BS}\ket{\text{MSQZ}} \notag\\
= & \bigotimes_{k\in\mathbb{N}}\ket{\text{EPR}}_{A,k;B,k}. 
\end{align}
Here we define the following transformation of annihilation operators $\hat{a}_{\Lambda,k} = (\hat{x}_{\Lambda,k}+i\hat{p}_{\Lambda,k})/\sqrt{2\hbar}$ as the beamsplitter interaction, 
\begin{align}
\hat{U}_\text{BS}
\begin{pmatrix}
\hat{a}_{A,k} \\ \hat{a}_{B,k}
\end{pmatrix}
\hat{U}_\text{BS}^\dagger = \frac{1}{\sqrt{2}}
\begin{pmatrix}
1 & -1 \\ 1 & 1
\end{pmatrix}
\begin{pmatrix}
\hat{a}_{A,k} \\ \hat{a}_{B,k}
\end{pmatrix}.
\end{align}
A unitary transformation of a state vector $\ket{\psi}\to\hat{U}\ket{\psi}$ is always associated with the corresponding transformation of nullifiers $\hat{N}\to\hat{U}\hat{N}\hat{U}^\dagger$. 
The nullifier representation of the multiplexed EPR state is 
\begin{subequations}
\begin{align}
(\hat{x}_{A,k}-\hat{x}_{B,k})\ket{\text{MEPR}} &= 0, \\
(\hat{p}_{A,k}+\hat{p}_{B,k})\ket{\text{MEPR}} &= 0. 
\end{align}
\end{subequations}

Then, they are staggered by the asymmetric delay, transforming the state vector as 
\begin{align}
\ket{\text{MEPR}} \to \ket{\text{MSEPR}}:= & \hat{U}_\text{DL}\ket{\text{MEPR}} \notag\\
= & \bigotimes_{k\in\mathbb{N}}\ket{\text{EPR}}_{A,k;B,k+1}. 
\end{align}
This is associated with the transformation of the nullifiers, 
\begin{subequations}
\begin{align}
(\hat{x}_{A,k}-\hat{x}_{B,k+1})\ket{\text{MSEPR}} &= 0, \\
(\hat{p}_{A,k}+\hat{p}_{B,k+1})\ket{\text{MSEPR}} &= 0.
\end{align}
\end{subequations}

Finally, another beamsplitter transformation leads to the ExEPR state, 
\begin{align}
\ket{\text{MSEPR}} \to \ket{\text{ExEPR}}:=\hat{U}_\text{BS}^\dagger\ket{\text{MSEPR}}, 
\end{align}
which is specified with the following nullifiers, 
\begin{subequations}
\begin{align}
(\hat{x}_{A,k}+\hat{x}_{B,k}+\hat{x}_{A,k+1}-\hat{x}_{B,k+1})\ket{\text{ExEPR}} &= 0, \\
(\hat{p}_{A,k}+\hat{p}_{B,k}-\hat{p}_{A,k+1}+\hat{p}_{B,k+1})\ket{\text{ExEPR}} &= 0.
\end{align}
\end{subequations}
We express these nullifiers of the ideal ExEPR state as $\hat{X}_k$ and $\hat{P}_k$.
\begin{subequations}
\label{eq:nullifiers}
\begin{align}
\hat{X}_k &:= \hat{x}_{A,k}+\hat{x}_{B,k}+\hat{x}_{A,k+1}-\hat{x}_{B,k+1}, \\
\hat{P}_k &:= \hat{p}_{A,k}+\hat{p}_{B,k}-\hat{p}_{A,k+1}+\hat{p}_{B,k+1}.
\end{align}
\end{subequations}

In fact, nullifiers span a vector space, and any linear combination of nullifiers is also a nullifier. 
The ExEPR state may also be specified with the following set of nullifiers, 
\begin{widetext}
\begin{subequations}
\begin{align}
\hat{X}_{A,2k}^\prime :=& \hat{X}_{2k-1}+\hat{X}_{2k} 
= 2\hat{x}_{A,2k}+\hat{x}_{A,2k-1}+\hat{x}_{B,2k-1}+\hat{x}_{A,2k+1}-\hat{x}_{B,2k+1}, \\
\hat{X}_{B,2k}^\prime :=& -\hat{X}_{2k-1}+\hat{X}_{2k} 
= 2\hat{x}_{B,2k}-\hat{x}_{A,2k-1}-\hat{x}_{B,2k-1}+\hat{x}_{A,2k+1}-\hat{x}_{B,2k+1}, \\
\hat{P}_{A,2k+1}^\prime :=& -\hat{P}_{2k}+\hat{P}_{2k+1} 
= 2\hat{p}_{A,2k+1}-\hat{p}_{A,2k}-\hat{p}_{B,2k}-\hat{p}_{A,2k+2}+\hat{p}_{B,2k+2}, \\
\hat{P}_{B,2k+1}^\prime :=& \hat{P}_{2k}+\hat{P}_{2k+1} 
= 2\hat{p}_{B,2k+1}+\hat{p}_{A,2k}+\hat{p}_{B,2k}-\hat{p}_{A,2k+2}+\hat{p}_{B,2k+2}.
\end{align}
\end{subequations}
They are equivalent to, via the redefinition of local phases $\hat{x}_{\Lambda,2k}\to\hat{p}_{\Lambda,2k}$, $\hat{p}_{\Lambda,2k}\to-\hat{x}_{\Lambda,2k}$, $\Lambda\in\{A,B\},k\in\mathbb{N}$, the nullifiers of a weighted CV cluster state~\cite{Yokoyama.nphoton2013}, 
\begin{subequations}
\begin{align}
\hat{H}_{A,k} :=& 2\hat{p}_{A,k}+\hat{x}_{A,k-1}+\hat{x}_{B,k-1}+\hat{x}_{A,k+1}-\hat{x}_{B,k+1}, \\
\hat{H}_{B,k} :=& 2\hat{p}_{B,k}-\hat{x}_{A,k-1}-\hat{x}_{B,k-1}+\hat{x}_{A,k+1}-\hat{x}_{B,k+1}.
\end{align}
\end{subequations}
\end{widetext}
These nullifiers explain the graph in (iv) of Fig.~\ref{fig:schematic}, where each edge expresses addition of the $\hat{x}$ quadrature operator to the $\hat{p}$ quadrature operator of the linked neighboring qumode to construct nullifiers. 
One can even use colors in the graph to classify nodes and edges~\cite{Menicucci.pra2011}.

\section{\label{sec:inseparability}Full-inseparability criteria}

\begin{figure*}[tb]
\centering
\includegraphics[clip]{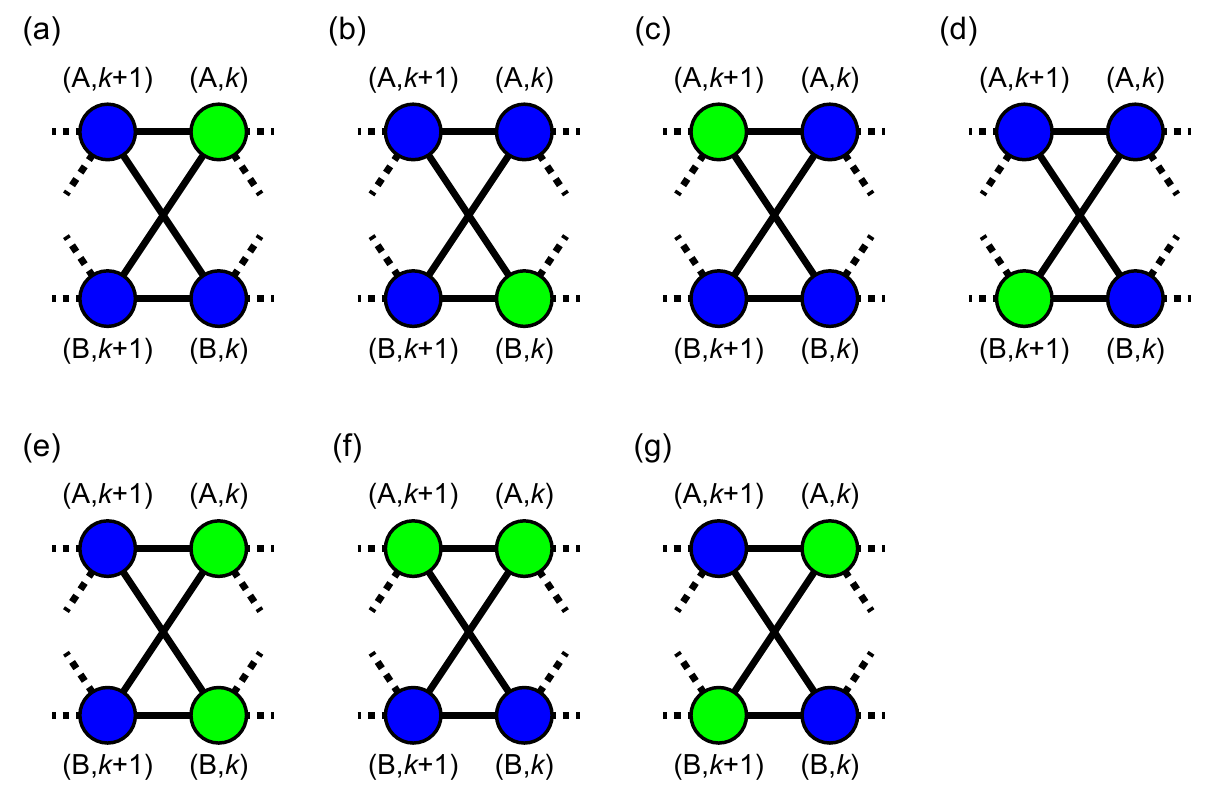}
\caption{
Seven patterns of dividing four qumodes into two parts. 
The colors of nodes, green and blue, represent the bipartition. 
}
\label{fig:bipartition}
\end{figure*}

Here we resort to the van Loock-Furusawa criterion for full inseparability~\cite{vanLoock.pra2003, Yokoyama.nphoton2013}. 
In the following, the set of all relevant qumodes is denoted by $S$, and we consider an arbitrary bipartition $\{S_\alpha,S_\beta\}$ of it. 
From the definition, the sets $S_\alpha$ and $S_\beta$ are nonempty subsets of $S$, satisfying $S_\beta = S \setminus S_\alpha$. 
Full inseparability means negation of separability for any bipartition $\{S_\alpha,S_\beta\}$ of the set $S$, i.e., the density operator never be in the form $\hat{\rho}_{S}=\sum_m p_m\hat{\rho}_{S_\alpha;m}\otimes\hat{\rho}_{S_\beta;m}$. 
Note that negation of separability for any bipartition is sufficient for negation of separability for any partition, consisting of more than two subsets.

The van Loock-Furusawa criterion is understood as translation of the positive partial transpose condition into the form of the uncertainty relation. 
The positive partial transpose condition is that, if a bipartite state is a separable state, i.e., the density operator is a mixture of product states $\hat{\rho}_{S}=\sum_{m}{p_m}\hat{\rho}_{S_{\alpha};m}\otimes\hat{\rho}_{S_{\beta};m}$, then the partially transposed state $\hat{\rho}_{S}^{T_{\beta}}=\sum_{m}{p_m}\hat{\rho}_{S_{\alpha};m}\otimes\hat{\rho}_{S_{\beta};m}^T$ must be also a physical state, represented by a positive semidefinite density operator. 
The contrapositive statement is that the bipartite state is inseparable (entangled) if the partially transposed state is unphysical, lacking positive semidefiniteness. 
A key insight is that the transposition of a density operator corresponds to mirror reflection $\hat{R}$ of the phase space, $\hat{R}(\hat{x},\hat{p})\hat{R}=(\hat{x},-\hat{p})$~\cite{Simon.prl2000}. 
We note $\hat{R}^2=\hat{1}$.

For an arbitrary pair of noncommutative observables $\hat{X}$ and $\hat{P}$, for which we suppose $\braket{\hat{X}}_{\hat{\rho}}=0$ and $\braket{\hat{P}}_{\hat{\rho}}=0$, the uncertainty relation is the inequality, 
\begin{align}
\braket{\hat{X}^2}_{\hat{\rho}}+\braket{\hat{P}^2}_{\hat{\rho}}
& \ge2\sqrt{\braket{\hat{X}^2}_{\hat{\rho}}}\sqrt{\braket{\hat{P}^2}_{\hat{\rho}}} \notag\\
& \ge|\braket{[\hat{X},\hat{P}]}_{\hat{\rho}}|, \label{eq:uncertainty}
\end{align}
where $\braket{\hat{O}}_{\hat{\rho}} := \Tr (\hat{O}\hat{\rho})$ is a mean value of an observable $\hat{O}$, explicitly showing the quantum state $\hat{\rho}$ for later use. 
Note that the first inequality is the inequality of arithmetic and geometric means. 

Here we consider observables which are linear combinations of quadrature operators, 
\begin{subequations}
\begin{align}
\hat{X}=\sum_{j\in S}c_j\hat{x}_j=(\sum_{j\in S_\alpha}c_j\hat{x}_j)+(\sum_{j\in S_\beta}c_j\hat{x}_j), \\
\hat{P}=\sum_{j\in S}d_j\hat{p}_j=(\sum_{j\in S_\alpha}d_j\hat{p}_j)+(\sum_{j\in S_\beta}d_j\hat{p}_j),
\end{align}
\end{subequations}
where the commutation relation of quadrature operators is $[\hat{x}_j,\hat{p}_{j^\prime}] = i\hbar\delta_{jj^\prime}$. 
Then, the uncertainty relation Eq.~\eqref{eq:uncertainty} is, 
\begin{align}
\braket{\hat{X}^2}_{\hat{\rho}_S}+\braket{\hat{P}^2}_{\hat{\rho}_S}
\ge \hbar|\sum_{j\in S_\alpha}c_jd_j+\sum_{j\in S_\beta}c_jd_j|. 
\end{align}
In addition, if the quantum state is separable with respect to the bipartition $\{S_{\alpha},S_{\beta}\}$, the uncertainty relation Eq.~\eqref{eq:uncertainty} is also valid for the partially transposed state~\cite{Toscano.pra2015}, 
\begin{align}
\braket{\hat{X}^2}_{\hat{\rho}_S}+\braket{\hat{P}^2}_{\hat{\rho}_S}
& = \braket{(\hat{R}_{\beta}\hat{X}\hat{R}_{\beta})^2}_{\hat{\rho}_S^{T_{\beta}}}+\braket{(\hat{R}_{\beta}\hat{P}\hat{R}_{\beta})^2}_{\hat{\rho}_S^{T_{\beta}}} \notag\\
& \ge \hbar|\sum_{j\in S_\alpha}c_jd_j-\sum_{j\in S_\beta}c_jd_j|. 
\end{align}
The above two inequalities are combined into an inequality, 
\begin{align}
\braket{\hat{X}^2}_{\hat{\rho}_S} + \braket{\hat{P}^2}_{\hat{\rho}_S} 
\ge \hbar\bigl(|\sum_{j\in S_\alpha}c_jd_j|+|\sum_{j\in S_\beta}c_jd_j|\bigr). 
\end{align}
If a quantum state $\hat{\rho}_S$ is separable with respect to the bipartition $\{S_\alpha,S_\beta\}$, this inequality is satisfied for any parameters $\{c_j,d_j\}_{j\in S}$. 
Contrapositively, if we could find some parameters $\{c_j,d_j\}_{j\in S}$ with which the inequality is broken, this is a witness of inseparability with respect to the bipartition $\{S_\alpha,S_\beta\}$. 
For full inseparability, we must consider a set of inequalities that denies separability for any bipartition of $S$.

We apply this criterion to the ExEPR state. 
The whole set of qumodes is $S=\{(\Lambda,k):\Lambda\in\{A,B\},k\in\mathbb{N}\}$. 
For any bipartition $\{S_\alpha,S_\beta\}$, there exists at least one subset consisting of four qumodes $S_k:=\{(A,k),(B,k),(A,k+1),(B,k+1)\}$ where a dividing line is running, i.e., the intersections $\{S_k\cap S_\alpha,S_k\cap S_\beta\}$ makes a partition of $S_k$. 
Then, if the separability of $S_k$ is negated, the separability of $S$ with respect to the bipartition $\{S_\alpha,S_\beta\}$ is also negated. 
This simplification is related to the fact that, if the whole density operator is separable as $\hat{\rho}_{S}=\sum_m p_m\hat{\rho}_{S_\alpha;m}\otimes\hat{\rho}_{S_\beta;m}$, the reduced density operator is also separable as 
\begin{align}
\hat{\rho}_{S_k}
= & \sum_m p_m(\Tr_{S_\alpha\setminus S_k}\hat{\rho}_{S_\alpha;m})\otimes(\Tr_{S_\beta\setminus S_k}\hat{\rho}_{S_\beta;m}) \notag\\
= & \sum_m p_m\hat{\rho}_{S_k\cap S_\alpha;m}\otimes\hat{\rho}_{S_k\cap S_\beta;m}. 
\end{align}
If full inseparability for all subset $S_k$ is proven, separability for any bipartition $\{S_\alpha,S_\beta\}$ is negated, which means full inseparability for the whole set $S$.

For the inseparabilities of the subset $S_k$, the residual variances of the nullifiers $\braket{\hat{X}_k^2}$ and $\braket{\hat{P}_k^2}$ are good candidates to construct inequalities, because they represent the strength of quantum correlations among qumodes. 
As depicted in Fig.~\ref{fig:bipartition}, bipartition $\{S_k\cap S_\alpha,S_k\cap S_\beta\}$ of a subset $S_k$ has seven patterns. 
We can construct individual inequalities for separability as below. 
\begin{subequations}
\begin{itemize}
\item[(a)] 
$S_k\cap S_\alpha = \{(A,k)\}$, \\ 
$S_k\cap S_\beta = \{(B,k),(A,k+1),(B,k+1)\}$, 
\begin{align}
\braket{\hat{X}_k^2}_{\hat{\rho}_S} + \braket{\hat{P}_k^2}_{\hat{\rho}_S} 
\geq \hbar\left(|1|+|1-1-1|\right) = 2\hbar.
\end{align}
\item[(b)] 
$S_k\cap S_\alpha = \{(B,k)\}$, \\ 
$S_k\cap S_\beta = \{(A,k),(A,k+1),(B,k+1)\}$, 
\begin{align}
\braket{\hat{X}_k^2}_{\hat{\rho}_S} + \braket{\hat{P}_k^2}_{\hat{\rho}_S} 
\geq \hbar\left(|1|+|1-1-1|\right) = 2\hbar.
\end{align}
\item[(c)] 
$S_k\cap S_\alpha = \{(A,k+1)\}$, \\
$S_k\cap S_\beta = \{(A,k),(B,k),(B,k+1)\}$, 
\begin{align}
\braket{\hat{X}_k^2}_{\hat{\rho}_S} + \braket{\hat{P}_k^2}_{\hat{\rho}_S} 
\geq \hbar\left(|-1|+|1+1-1|\right) = 2\hbar.
\end{align}
\item[(d)] 
$S_k\cap S_\alpha = \{(B,k+1)\}$, \\
$S_k\cap S_\beta = \{(A,k),(B,k),(A,k+1)\}$, 
\begin{align}
\braket{\hat{X}_k^2}_{\hat{\rho}_S} + \braket{\hat{P}_k^2}_{\hat{\rho}_S} 
\geq \hbar\left(|-1|+|1+1-1|\right) = 2\hbar.
\end{align}
\item[(e)] 
$S_k\cap S_\alpha = \{(A,k),(B,k)\}$, \\
$S_k\cap S_\beta = \{(A,k+1),(B,k+1)\}$,  
\begin{align}
\braket{\hat{X}_k^2}_{\hat{\rho}_S} + \braket{\hat{P}_k^2}_{\hat{\rho}_S} 
\geq \hbar\left(|1+1|+|-1-1|\right) = 4\hbar.
\end{align}
\item[(f)] 
$S_k\cap S_\alpha = \{(A,k),(A,k+1)\}$, \\
$S_k\cap S_\beta = \{(B,k),(B,k+1)\}$, 
\begin{align}
\braket{\hat{X}_k^2}_{\hat{\rho}_S} + \braket{\hat{P}_{k+1}^2}_{\hat{\rho}_S} 
\geq \hbar\left(|1|+|-1|\right) = 2\hbar.
\end{align}
\item[(g)] 
$S_k\cap S_\alpha = \{(A,k),(B,k+1)\}$, \\
$S_k\cap S_\beta = \{(B,k),(A,k+1)\}$, 
\begin{align}
\braket{\hat{X}_k^2}_{\hat{\rho}_S} + \braket{\hat{P}_{k+1}^2}_{\hat{\rho}_S} 
\geq \hbar\left(|-1|+|1|\right) = 2\hbar.
\end{align}
\end{itemize}
\end{subequations}
Note that $\braket{\hat{P}_{k+1}^2}$ is utilized in the inequalities instead of $\braket{\hat{P}_k^2}$ for the last two patterns. 
Considering the above seven patterns for all $k\in\mathbb{N}$, we can see that the following set of inequalities are a sufficient condition for full inseparability. 
\begin{align}
\braket{\hat{X}_k^2} &< \hbar, & \braket{\hat{P}_k^2} &< \hbar, & \text{for all $k$}. 
\end{align}
Comparing them with Eq.~\eqref{eq:nullifier_var}, these inequalities corresponds to $e^{-2r_{x,k}}<1/2$ and $e^{-2r_{p,k}}<1/2$, which means that more than $-$3~dB of squeezing for all nullifiers are sufficient to verify full inseparability.

\end{document}